\documentclass[twoside,twocolumn,fleqn,superscriptaddress,showkeys]{revtex4-2}
\usepackage{xcolor}    
\usepackage{graphicx}  
\usepackage{amssymb}   
\usepackage{amsmath}   
\usepackage{bm}        
\usepackage{times}     
\usepackage{fancyhdr}  
\usepackage{lineno}
\usepackage{color}
\newcommand{\fix}[1]{\textcolor{black}{#1}}

\begin{document} 

\title{Calibration of detector time constant with a thermal source for the  \textsc{Polarbear}-2A CMB polarization experiment} 

\author{S. Takatori} 
\email[Corresponding author: ]{takatori@okayama-u.ac.jp}
\affiliation{Research Institute for Interdisciplinary Science (RIIS), Okayama University, Tsushima-naka 3-1-1 Kita-ku, Okayama 700-8530, Japan}

\author{M. Hasegawa}
\affiliation{International Center for Quantum-field Measurement Systems for Studies of the Universe and Particles (QUP), High Energy Accelerator Research Organization (KEK), Tsukuba, Ibaraki 305-0801, Japan}
\affiliation{Institute of Particle and Nuclear Studies (IPNS), High Energy Accelerator Research Organization (KEK), 1-1 Oho, Tsukuba, Ibaraki 305-0801, Japan}

\author{M. Hazumi}

\affiliation{International Center for Quantum-field Measurement Systems for Studies of the Universe and Particles (QUP), High Energy Accelerator Research Organization (KEK), Tsukuba, Ibaraki 305-0801, Japan}
\affiliation{Institute of Particle and Nuclear Studies (IPNS), High Energy Accelerator Research Organization (KEK), 1-1 Oho, Tsukuba, Ibaraki 305-0801, Japan}
\affiliation{Institute of Space and Astronautical Science, Japan Aerospace Exploration Agency (JAXA), Sagamihara 252-5210, Japan}
\affiliation{Kavli Institute for the Physics and Mathematics of the Universe (Kavli IPMU), University of Tokyo
5-1-5 Kashiwa-no-Ha, Kashiwa City, Chiba 277-8583, Japan}

\author{D. Kaneko}
\affiliation{International Center for Quantum-field Measurement Systems for Studies of the Universe and Particles (QUP), High Energy Accelerator Research Organization (KEK), Tsukuba, Ibaraki 305-0801, Japan}

\author{N. Katayama}
\affiliation{Kavli Institute for the Physics and Mathematics of the Universe (Kavli IPMU), University of Tokyo 5-1-5 Kashiwa-no-Ha, Kashiwa City, Chiba 277-8583, Japan}%

\author{A. T. Lee}
\affiliation{Department of Physics, University of California, Berkeley, CA 94720, USA}
\affiliation{Physics Division, Lawrence Berkeley National Laboratory, Berkeley, CA 94720, USA}

\author{S. Takakura}
\affiliation{Department of Physics, Faculty of Science, Kyoto University, Kyoto, Kyoto 606-8502, Japan}

\author{T. Tomaru}
\affiliation{Gravitational Wave Project Office, National Astronomical Observatory of Japan (NAOJ), Mitaka City, Tokyo 181-8588, Japan}

\author{T. Adkins}
\affiliation{Department of Physics, University of California, Berkeley, CA 94720, USA}

\author{D. Barron}
\affiliation{Department of Physics and Astronomy, University of New Mexico, Albuquerque, NM 87131, USA}

\author{Y. Chinone}
\affiliation{International Center for Quantum-field Measurement Systems for Studies of the Universe and Particles (QUP), High Energy Accelerator Research Organization (KEK), Tsukuba, Ibaraki 305-0801, Japan}
\affiliation{Institute of Particle and Nuclear Studies (IPNS), High Energy Accelerator Research Organization (KEK), 1-1 Oho, Tsukuba, Ibaraki 305-0801, Japan}
\affiliation{Kavli Institute for the Physics and Mathematics of the Universe (Kavli IPMU), University of Tokyo 5-1-5 Kashiwa-no-Ha, Kashiwa City, Chiba 277-8583, Japan}

\author{K. T. Crowley}
\affiliation{Department of Physics, University of California, Berkeley, CA 94720, USA}

\author{T. de Haan}
\affiliation{Institute of Particle and Nuclear Studies (IPNS), High Energy Accelerator Research Organization (KEK), 1-1 Oho, Tsukuba, Ibaraki 305-0801, Japan}
\affiliation{International Center for Quantum-field Measurement Systems for Studies of the Universe and Particles (QUP), High Energy Accelerator Research Organization (KEK), Tsukuba, Ibaraki 305-0801, Japan}

\author{T. Elleflot}
\affiliation{Physics Division, Lawrence Berkeley National Laboratory, Berkeley, CA 94720, USA}

\author{N. Farias}
\affiliation{Department of Physics, University of California, Berkeley, CA 94720, USA}

\author{C. Feng}
\affiliation{University of Science and Technology of China, 230052, PRC}

\author{T. Fujino}
\affiliation{Yokohama National University, Yokohama, Kanagawa 240-8501, Japan}

\author{J. C. Groh}
\affiliation{Superconductive Electronics Group, National Institute of Standards and Technology, Boulder, CO 80305, USA}

\author{H. Hirose}
\affiliation{Yokohama National University, Yokohama, Kanagawa 240-8501, Japan}

\author{F. Matsuda}
\affiliation{Institute of Space and Astronautical Science, Japan Aerospace Exploration Agency (JAXA), Sagamihara 252-5210, Japan}

\author{H. Nishino}
\affiliation{Japan Synchrotron Radiation Research Institute (JASRI), Sayo-gun, Hyogo 679-5198, Japan}%

\author{Y. Segawa}
\affiliation{School of High Energy Accelerator Science, The Graduate University for Advanced Studies, SOKENDAI, Kanagawa 240-0193, Japan}

\author{P. Siritanasak}
\affiliation{National Astronomical Research Institute of Thailand, Chiangmai 50180, Thailand}

\author{A. Suzuki}
\affiliation{Physics Division, Lawrence Berkeley National Laboratory, Berkeley, CA 94720, USA}

\author{K. Yamada}
\affiliation{Department of Physics, The University of Tokyo, Tokyo 113-0033, Japan}


\date{\today}

\begin{abstract}
The Simons Array (SA) project is a ground-based Cosmic Microwave Background (CMB) polarization experiment. 
The SA observes the sky \fix{using} three telescopes, and \textsc{Polarbear}-2A (PB-2A) is the receiver system on the first telescope. . 
\fix{For the ground-based experiment,} atmospheric fluctuation is the primary noise source \fix{that} could cause polarization leakage. In the PB-2A receiver system, a continuously rotating \fix{half-wave} plate (HWP) is used to mitigate the polarization leakage. \fix{However}, due to the rapid modulation of the polarization signal, the uncertainty in the time constant of the detector results in an uncertainty in the polarization angle. For PB-2A, the time constant of each bolometer needs to be calibrated at the sub-millisecond level \fix{to avoid introducing} bias to the polarization signal. 
We have developed a new calibrator system \fix{that} can be used to calibrate the time constants of the detectors.  
In this study, we present the design of the calibration system 
and the preliminary results of the time constant calibration for PB-2A.
\end{abstract}


\maketitle  
\thispagestyle{fancy}  


\section{INTRODUCTION} 
Precise measurements of Cosmic Microwave Background (CMB) polarization, especially detecting the odd-parity CMB polarization pattern -- 
called the B-mode -- is a unique and powerful way of exploring the early universe. The Simons Array (SA) project \cite{SA_N_Stebor} is a ground-based CMB polarization experiment \fix{conducted} in the Atacama desert, Chile, at an altitude of 5,200 meters. 
SA aims \fix{to characterize} B-mode signals from inflationary gravitational waves and gravitational lensing effects. 
The projected constraint on a tensor-to-scalar ratio $r$ is at the level of $\sigma(r)$ = 0.006 by the inflationary B-modes, and the sensitivity to the sum of neutrino masses is 40 meV (68\% C.L.) by the lensing B-modes \cite{Suzuki_2016}. 
 To improve the sensitivity and separation of foreground radiation, SA uses a total of over 20,000 Transition Edge Sensor (TES) bolometers \cite{Irwin2005} in four frequency bands (90, 150, 220, and 280 GHz). 
SA observes the sky using three telescopes.  \textsc{Polarbear}-2A (PB-2A) \cite{Hasegawa_2018} is the receiver system on the first of three telescopes, with 7588 TES bolometers in two observing bands of 90 and 150 GHz. 
PB-2A had its first light in 2019 \cite{Kaneko_2020} and is currently being used for observing the CMB. 

In ground-based CMB experiments, the sensitivity is limited by \fix{low-frequency} $1/f$ noise especially for observing large \fix{angular-scale} B-modes caused by inflationary gravitational waves.
In PB-2A, a \fix{continuously rotating} half-wave plate (HWP) \cite{Hill_2016} is used as a polarization modulator in front of the receiver window to modulate the polarization signal and reduce the effect of the $1/f$ noise \cite{Takakura_2017}. 
In modulating the polarization signal, the uncertainty of the detector's time constant (time delay in response) causes uncertainties in the observed polarization angle, further resulting in the appearance of a pseudo B-mode \cite{Shimon_2008}.

To calibrate the time constant and the relative gain of the detector, we use the blackbody emission from a thermal source that can cover two observing bands simultaneously as a reference source. 
In this study, we explore the new calibration system using this thermal source 
and \fix{demonstrate the measurement} of the time constant during the initial site commissioning test of PB-2A.

\section{system design} 

TES bolometers were `tuned' around the transition temperature before starting the daily observations. The time constant of each bolometer is dependent on the tuned status, which is affected by 
the focal plane temperature and loading (sky) conditions. 
Therefore, the time constant $\tau$ needs to be calibrated \fix{for} each observation. 
In PB-2A, the HWP was continuously rotated at 2 Hz in front of the receiver to modulate the polarization signal of the incoming CMB light and shifted it to a higher frequency to avoid the effect of $1/f$ noise \cite{Hill_2016}. 
When calibrating the time constants, the uncertainty in the time constant led to uncertainty in the HWP angle for each observation, and ultimately, to variations in the absolute polarization angle.

The uncertainty of the measured time constant of the TES bolometer $\delta \tau$ became the uncertainty  $\delta \theta$ in the polarization angle 
$\theta$. If the rotational speed of the HWP was $f_{\rm HWP}$ Hz, then the \fix{uncertainty in the polarization angle} would be $\delta \theta  \approx 2 \cdot 2 \pi f_{\rm HWP} \delta \tau $  \cite{tau_pol_angle}.  
PB-2A calibrated the absolute polarization angles by observing a polarized celestial source such as tau~A. 
We assumed that the uncertainty in the absolute polarization angle \fix{may be attributed solely to} the variation of polarization angle due to the uncertainty in the time constant. 
The polarization angle uncertainty $\delta \theta$ \fix{should} be smaller than $0.1^\circ$ \cite{Marty_pol_angle}, so that the following equation holds.  

\begin{equation}
\delta \tau < \frac{\delta \theta}{2 \cdot 2 \pi f_{\rm HWP} } \approx 0.069 {\rm ms}\Bigl(\frac{\delta \theta}{0.1^\circ}\Bigr)\Bigl(\frac{f_{\rm HWP}}{2\,\mathrm{Hz}}\Bigr)^{-1}.
\end{equation}
The uncertainties in the time constant should be less than 0.07 ms in each observation.

A TES output $I$ is modeled using a one-pole model as  
\begin{equation}
I(f) =\frac{I_0}{\sqrt{1+(2\pi f \tau)^2}}, 
 \label{eq:onepole}
\end{equation}
where $f$ is a signal modulation frequency, \fix{$\tau$ is the time constant}  and $I_0 $ is the signal amplitude. The expected time constant for PB-2A ranged from 1 to 5 ms \cite{Darcy_PB2_readout}. 
To simplify the discussion,  
\fix{the uncertainty of the time constant may be defined as}
\begin{equation}
\delta {\tau} \equiv \left(\frac{dI}{d\tau}\right)^{-1} \delta I,
\label{eq:sigma_tau}
\end{equation}
\fix{where $\delta I$ is uncertainty of the signal amplitude. }
From \fix{Eq.~\ref{eq:onepole}}, the estimated uncertainty of the time constant is  
\fix{
\begin{equation} \label{eq:tauresolution}
\delta \tau
= \frac{ (1+(2\pi\tau f_{\mathrm{max}})^2)^{\frac{3}{2}}}{(2\pi f_{\mathrm{max}})^2 \tau} \frac{\delta I}{I_0}.
\end{equation}
}
For example, when $\fix{\delta I/I_0} = o(10^{-3})$,  $\tau = 1~\rm{ms}$ and $f_{\mathrm{max}} = \fix{44}~\rm{Hz}$, then $\delta \tau \fix{\sim 0.01} ~\rm{ms}$, which is small for our requirement. 

\begin{figure}[htbp]
  \begin{center}
  \includegraphics[width=70mm]{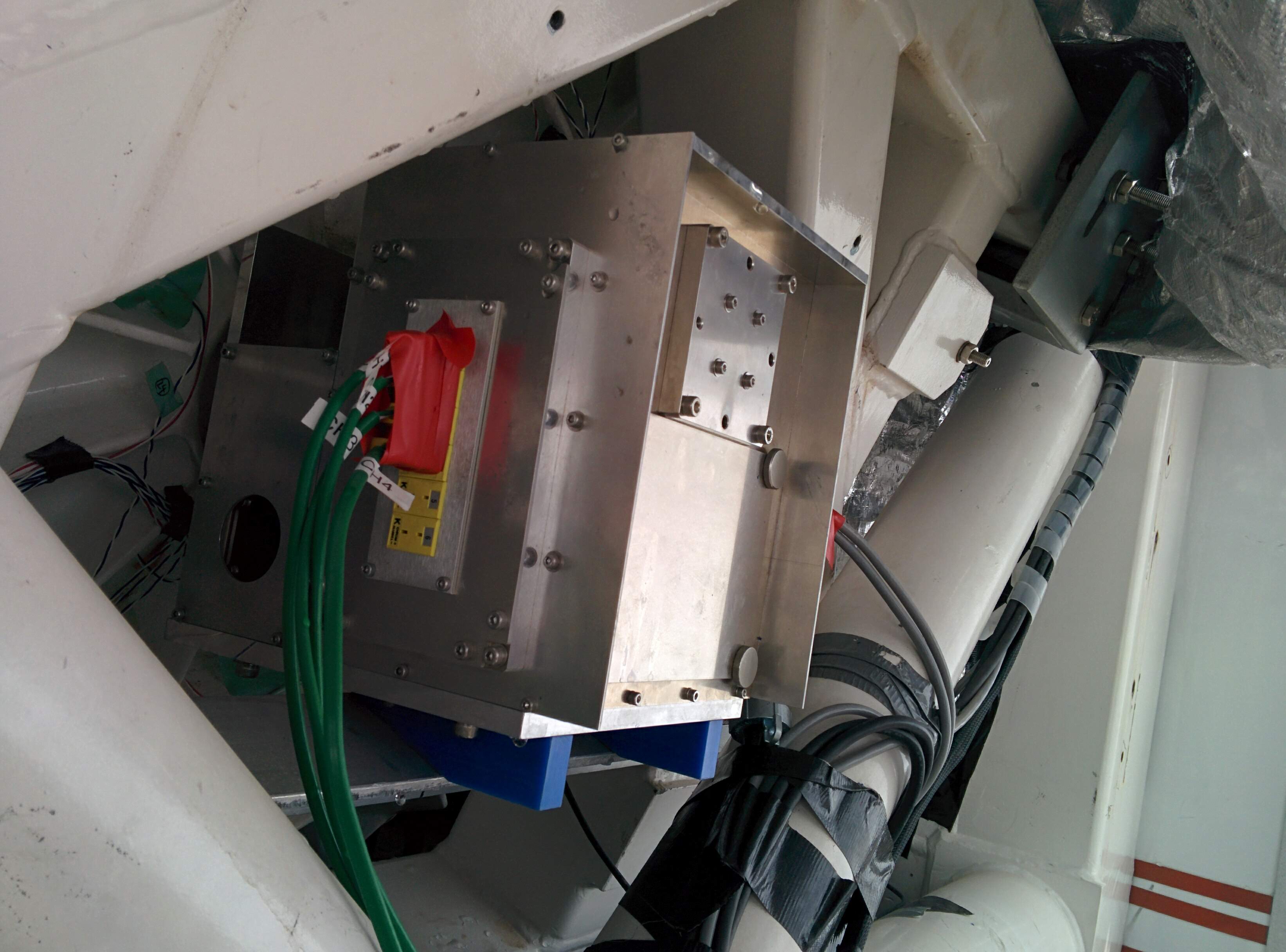}
 \end{center}
 \caption{\fix{Experimental system mounted behind secondary mirror.} It emits a reference signal through a light-pipe in the 9-mm hole of the secondary mirror. }
 \label{fig:mount}
 \end{figure}

The system, including the thermal source (1000 K), was mounted behind the telescope's secondary mirror, as shown in FIG. \ref{fig:mount}. 
The radiation was emitted through a 9-mm hole at the center of the secondary mirror through a light pipe into the receiver.
Using the expected noise-equivalent temperature value, $\mathrm{NET} = 360~\rm{\mu K \sqrt{s}}$ \cite{Suzuki_2016}, the effective temperature of the thermal source $ T_{\rm s}  \sim 0.04 $ K and the measurement time $t_{\rm mes}=$ 120 s,  $\fix{\delta I/I_0}$ can be estimated as $\fix{\delta I/I_0} = \mathrm{NET}/T_{\rm s}\sqrt{t_{\rm mes}} \sim 8 \times 10^{-4}$.

 \begin{figure}[htbp]
 \begin{center}
 \includegraphics[width=78mm]{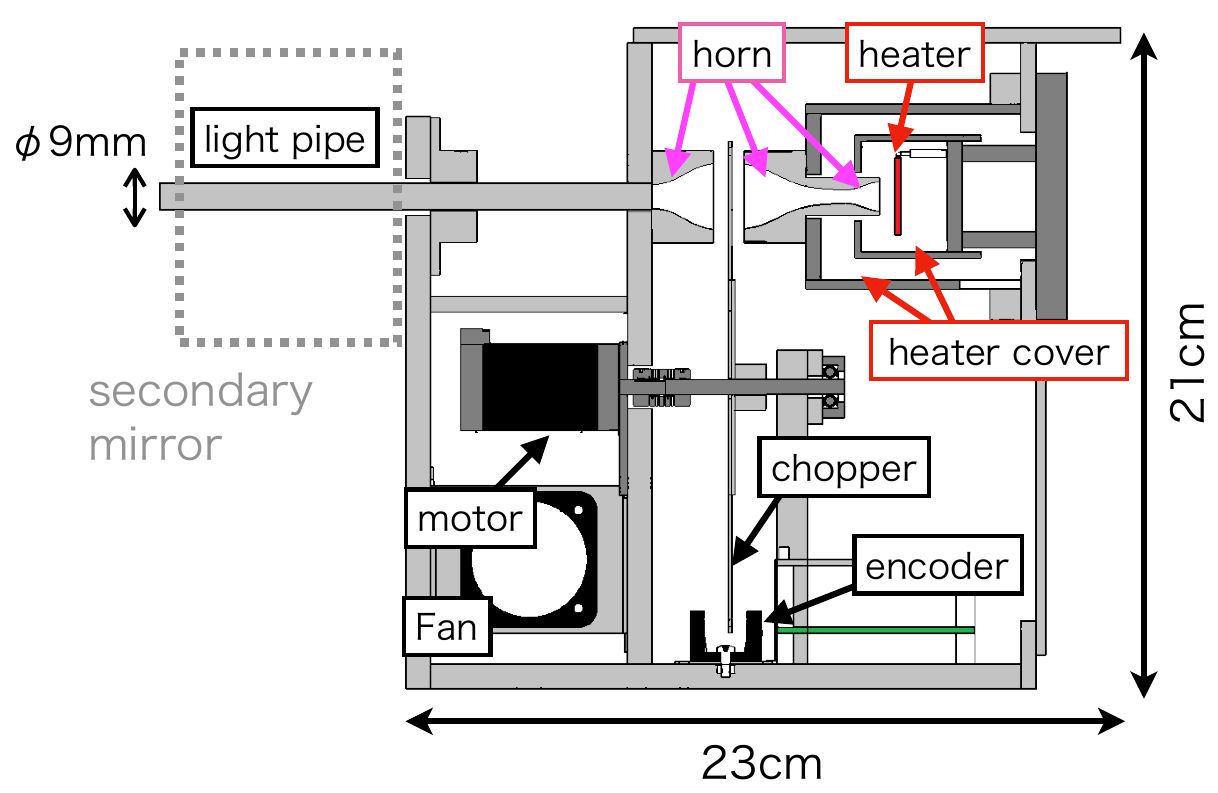}
 \end{center}
  \caption{Side view of the system. System has a small chopped thermal source, where a ceramic heater is used as a blackbody radiator to cover up observation frequency bands.}
 \label{fig:one}
\end{figure}

\section{\label{sec:level2}Overview of Calibration system}
A side view of the calibration system is shown in FIG. \ref{fig:one}.  The entire system is a small box, approximately in the shape of a 20-cm cube, to be mounted in the limited space behind the telescope's secondary mirror, as shown in FIG. \ref{fig:mount}.
The system employs a 25\,mm $\times$ 25\,mm ceramic heater (MS-1000 \cite{sakaguchi-dennetsu}) with a monitored temperature of approximately 1000\,K, as the blackbody radiator. 
The nominal operating voltage of the heater is set as 40 V AC and the wattage of the heater is approximately 30 W. 
A thermocouple is attached with an alumina bond to monitor the heater temperature at 10-second intervals. 
The RMS value of the heater temperature stability over the entire season is 2 K, \fix{while it} is 1 K after \fix{the seasonal variations are eliminated}.

The system employs Winston cone horns, a type of compound parabolic concentrator (CPC) \cite{Winston:70}, to cover wide frequency bands.  
The horns are used to collect radiation from the heater surface and to improve the coupling efficiency before and after the optical chopper wheel.
The \fix{geometry} of the horn is designed to efficiently focus light into a light pipe with an inner diameter of 8 mm, which penetrates through the secondary mirror. 
The heater source section has a double-layered cover for improving the temperature stability. 
In the case of a heater breakdown under the harsh environment of the observation site, the first layer of the light source section alone can be replaced from behind \fix{in the form of a cartridge} without \fix{dismantling} the entire system. 
The system has remained operational throughout the past three years without any breakdowns.

The radiation from the heater is modulated by a rotating six-blade chopper to measure the time delay in the response of the detecter. 
An encoder (photosensor, EE-SX1140) is mounted on the opposite side of the light pipe to monitor the timing of the opening and closing of the chopper blade. 
The signal from the encoder is synchronized with the time-ordered bolometer data, whose sampling rate is 152\,Hz \cite{nishino2020} for analysis. 
The rotational stability of the chopper is typically 0.2\%. It exhibits a good chopping stability, 
and the chopping frequency increases to the Nyquist frequency of the sample rate. 
The stepping motor for the rotating chopper, which needs to be cooled by fans, is placed in a separate room to improve the internal temperature stability of the box. 

\section{Site commissioning test} 

The PB-2A receiver and the developed thermal source system were installed on the telescope in late 2018 and early 2019. PB-2A had its first light in 2019 \fix{for a} 
 demonstration that took place during the early commissioning test in 2019. 
The number of biased bolometers for which the signal from this system appears to have a signal-to-noise ratio of 20 or better is approximately 99\,\% across the entire focal plane of the PB-2A receiver. 
The time constant of the bolometers is measured by checking the detector time response to the modulated signal from this system. 
The rotational frequencies of the chopper are 5, 9, 13, 19, 29, 37, and 44 Hz, and the signal is observed for 120 s each, for a total of approximately 14 min. 
Before data analysis, the time-ordered  bolometer data was passed through high-pass filters and 
 the mean of the baseline was subtracted. 
In data analysis, we chunked and stacked the time-ordered bolometer data according to the encoder phase. Then, the amplitude of the modulated signal from the system was estimated by fitting the stacked data. We used the fitting function of the sum of the sinusoidal wave, $A(\theta)=\sum_{n=1}^{7}A_n \sin(n\theta + \phi_n) ~ [0 \leqq \theta < 2 \pi]$,   $A_1$ was assumed as an amplitude of the modulated signal.
The one-pole time constant model (Eq.~\ref{eq:onepole}) was used to fit the point of the signal amplitude at each frequency.
An example of the fitting is shown in FIG. \ref{fig:four}. 
At the same time, the relative gain $I_0$ of the detector can \fix{also be obtained from the fit.}

\begin{figure}[htbp]
 \begin{center}
  \includegraphics[width=65mm]{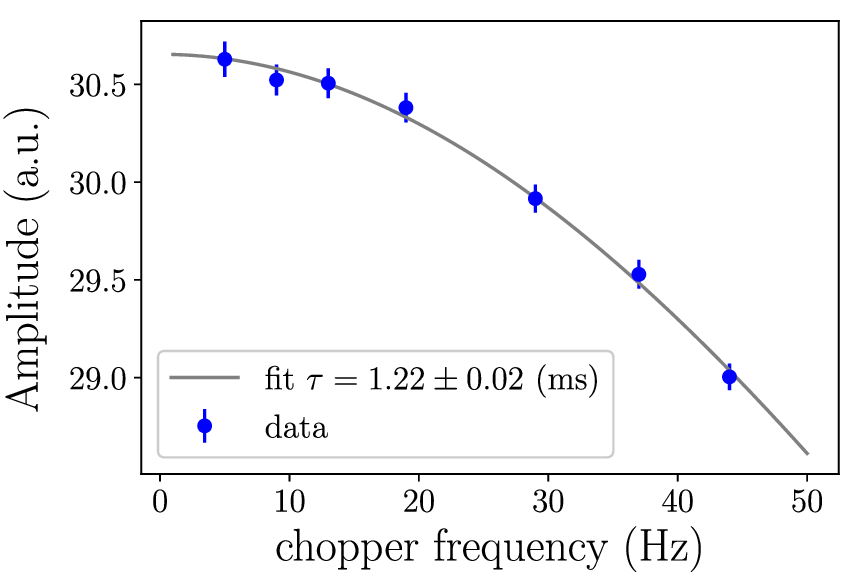}
  \end{center}
 \caption{
 Example of signal amplitudes from one calibration sequence in a 150 GHz bolometer. The time constant of the bolometer is measured by fitting the point of the signal amplitude at each frequency with the one-pole model.}
 \label{fig:four}
 \setcounter{figure}{4}
 \end{figure}
 
\fix{FIG.  \ref{fig:tau_err}} shows a scatter plot of the time constants and their uncertainties obtained by fitting the response of each bolometer. 
The signal-to-noise ratio based on the relative gain $I_0$ obtained by fitting is plotted using a color scale. 
As shown in Eq.~\ref{eq:tauresolution}, the magnitude of the time constant uncertainties is proportional to the noise level, and the trend of the experimental data \fix{is in accordance with the equation.} 
The bolometer with the highest signal-to-noise ratio meets the time constant calibration accuracy requirement described in the above section. If the noise level is close to the design value, Eq.~\ref{eq:tauresolution} suggests that, by improving the noise level, the uncertainty in the time constant can be reduced and the required calibration accuracy can be achieved. As this test was performed at the beginning of the commissioning test, the time constants were slower and noise level were higher than expected because the bias tuning of the bolometers was not optimized. 

\begin{figure*}[t]
\setcounter{figure}{3}
 \begin{center}
  \includegraphics[width=14cm]{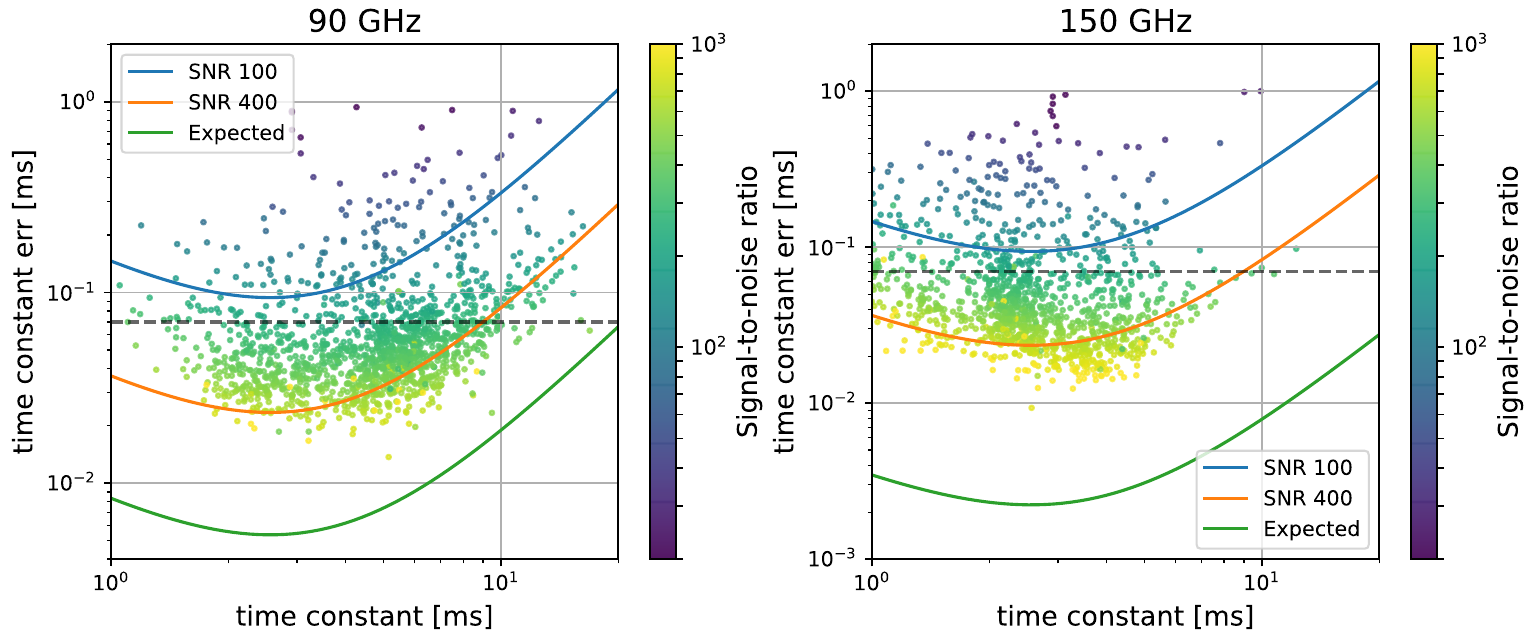}
  \end{center}
 \caption{Scatter plot of the time constants and their uncertainties. The signal-to-noise ratio of the signal from the thermal source is plotted using a color scale. The blue, orange, and green lines show \fix{$\delta \tau$} from Eq.~\ref{eq:tauresolution} with signal-to-noise ratio, \fix{$I_0/\delta I$} , of 100, 400, and that for the expected sensitivity, respectively. 
The bolometer with the higher signal-to-noise ratio meets the calibration accuracy requirement ($<$ 0.07 ms) in the expected noise level. 
This data was taken in an initial commissioning phase, and sufficient calibration accuracy can be achieved by improving the expected noise level ($\mathrm{NET} = 360 \rm{\mu K \sqrt{s}}$).}
 \label{fig:tau_err} 
\end{figure*}

\begin{figure}[htbp]
 \begin{center}
 \includegraphics[width=65mm]{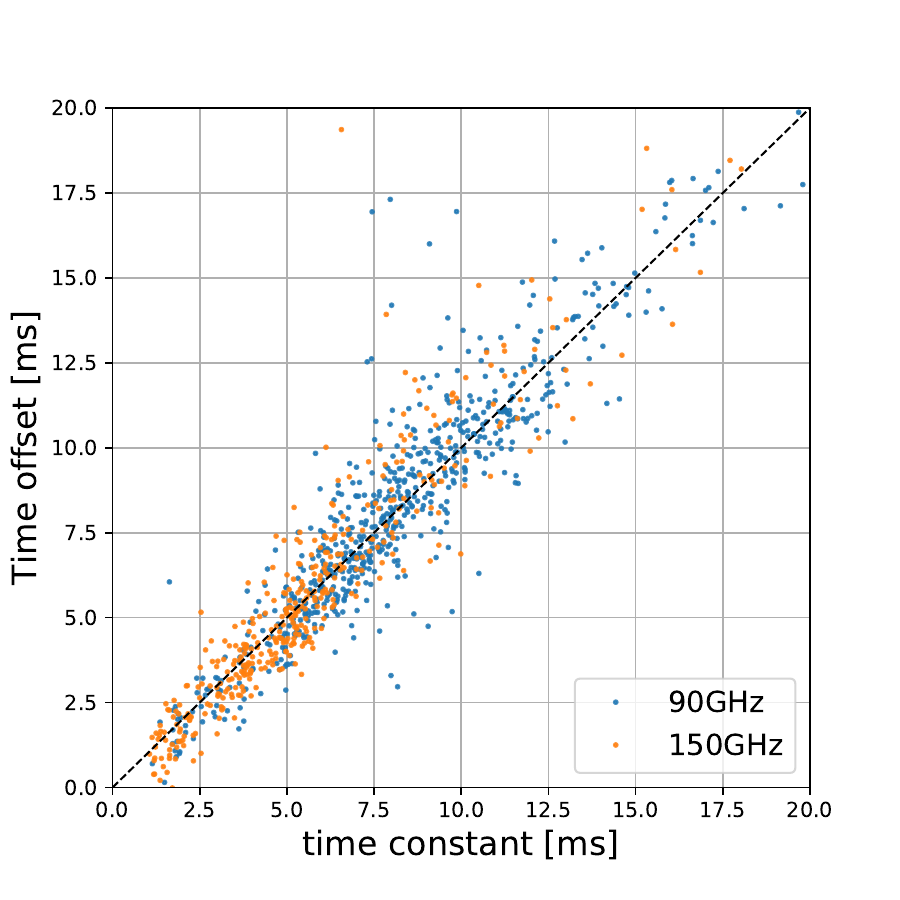}
 \end{center}
 \caption{Comparison of time constants between calibration system and astronomical point source. This scatter plot shows a strong correlation. }
 \label{fig:planet_tau}
\end{figure}

The time constant was also obtained from the data of the astronomical point source (Jupiter) observation during commissioning and compared with the time constant obtained using this calibration system. 
In SA, the telescope observes the astronomical point source by changing the elevation slightly and scanning 
left and right in the azimuth direction. 
The effect of the time constant appears as a shift in the apparent position along the scanning direction. 
The position of the celestial body was obtained by fitting with a 2D Gaussian. The time constant value obtained from the difference in the position of point sources is shown in FIG. \ref{fig:planet_tau}. 
We checked the consistency of the time constants obtained from our calibration system and from other calibration point sources. 
Note that as this data was acquired during the ongoing alignment of the receiver position, the variation in time offset was not directly indicative of the precision of the time constant measurement using the astronomical point source.

\section{CONCLUSIONS} 
In the PB-2A receiver system, a continuously rotating HWP was used to mitigate the low-frequency noise. With rapid polarization modulation, the uncertainty in the time constant of the detector resulted in an uncertainty in the polarization angle. 
To calibrate the time constant of the detector, we developed a new calibration system with a thermal source. The system employed a thermal ceramic heater (1000 K) as a blackbody radiator to cover the entire frequency band of the PB-2A observation.
In early commissioning tests at the observation site in Chile in 2019, we demonstrated that the time constant could be measured for several thousand detectors simultaneously and confirmed that our system could calibrate each bolometer at a sub-ms level. 
This system will contribute to the calibration of the detector time constant for PB-2A observation.

\begin{acknowledgments}
This work was supported by the Grant-in-Aid for JSPS Fellows (Grant Number JP19J10652),  JSPS KAKENHI Grant Number JP22K20371, Ozaki exchange program 2019 and SOKENDAI Advanced Research Course program. The financial support of the organization for diversity management, Okayama University is also acknowledged. 
S.T. acknowledges the support by Okayama University Dispatch Project for Female Faculties.
\fix{M.H. acknowledges the support from World Premier International Research Center Initiative (WPI,) MEXT, Japan and JSPS KAKENHI Grant Number 22H04945.}
I would like to thank T. Hamada \fix{and S. Kumazaki} for their helpful comments and discussions.
\end{acknowledgments}

\bibliography{aip-cp-samp_v2}
\bibliographystyle{unsrt}



\end{document}